\documentclass[
twocolumn,
]{ceurart}

\usepackage{lmodern}

\begin{document}

\copyrightyear{2021}
\copyrightclause{Copyright for this paper by its authors.
  Use permitted under Creative Commons License Attribution 4.0
  International (CC BY 4.0).}

\conference{ACM International Conference on Intelligent
Virtual Agents (IVA ’24), September 16–19, 2024, Glasgow, United Kingdom}

\title{Exploring the role of embodiment on intimacy perception in a multiparty collaborative task}

\author[1]{Amine Benamara}[%
orcid=0000-0003-2104-9535,
email=benamara@lisn.upsaclay.fr
]
\author[1]{Céline Clavel}[%
orcid=0000-0002-2253-7963,
email=clavel@lisn.upsaclay.fr
]

\author[1]{Brian Ravenet}[%
orcid=0000-0001-6824-4800,
email=ravenet@lisn.upsaclay.fr
]

\author[1]{Nicolas Sabouret}[%
orcid=0000-0002-7458-6732,
email=sabouret@lisn.upsaclay.fr
]

\author[2]{Julien Saunier}[%
orcid=0000-0002-7385-4395,
email=julien.saunier@insa-rouen.fr
]
\address[1]{Université Paris-Saclay, CNRS, Laboratoire Interdisciplinaire des Sciences du Numérique}
\address[2]{INSA Rouen Normandie, Univ Rouen Normandie, Université Le Havre
Normandie, Normandie Univ, LITIS UR 4108, F-76000 Rouen, France}

\begin{abstract}
During collaborative board games, cohesion represents a key aspect to define a well functionning group. From the success of the task to the developement of interpersonal relationship, this concept covers many aspects of group dynamics. The goal of our work is to investigate the factors that impact cohesion in a group, and specifically the relevant social skills that improve collaboration between multiple entities. In this article, we focus on the role of embodiement on different aspects of an interaction. We propose an experimental protocol, based on a collected corpus of humans playing a collaborative board game, to study how different agents' embodiment affect the perception of these agents and of the group as a whole. We conclude by presenting an outline of the problematics of the conception of the protocol and of multi-agent system related challenges.
\end{abstract}


\maketitle

\section{Introduction}

Small-group interactions in the context of a collaborative task represent an important part of everyday activities, like furniture assembly with a relative, work meetings or collaborative board games. These interactions require mutual cooperation between group members to achieve a common goal. Studies in social science and psychology investigated the different factors that affect group dynamics, and thus the success of a collaborative task, from a group-level and from an individual-level perspective \citep{Butera2019, Abrams2020}. On the group-level, factors as composition, spatial configuration or structure of the group have a known impact on the group dynamics \citep{Oliveira2021}. Researchers also focus on concepts such as entitativity, the outside perception of the group, or more frequently cohesion, which refers to the forces that hold a group together \citep{Gillet2022}.

The concept of cohesion, often linked to group performance and viability, received particular attention from researchers \citep{Severt2015}. Despite a large variety of definitions and models, they mostly agree on two interrelated dimensions of cohesion \citep{Maman2022}: task cohesion, which refers to the engagement of the group towards the task, and social cohesion, focused on the interpersonal relations between the group's members. On the individual level, we find the same distinction between task-related features, like the ability of each member to summarize information during work meeting, or a good understanding of the games' rules during a collaborative board game, and social-related features, like the ability to create social bonds with other members, the member's perceived inclusion in the group, or attitudes and emotions displayed by each member \citep{Maman2022}.

These studies on group processes also highlighted the bidirectional dynamics between the individual and the group : each individual impacts the group processes, while the group impacts each individual's behaviors \citep{Forsyth2010}. This implies that the introduction of controlled agents in a group of humans could lead to a positive impact on the group dynamics. With the recent development of socially interactive agents and robots, researchers have tried to figure out how these entities could be integrated to small group of humans, and the effect they can have on individuals and group processes \citep{Koutsombogera2019}. They found that the number of agents \citep{Fraune2015}, the roles of the agents during the interaction \citep{Short2017}, the agents' non-verbal behaviors \citep{DeConinck2019, Kantharaju2020, Kubasova2019, Tarnec2023} and their embodiments \citep{Shamekhi2018} can have a positive influence on the group dynamics. Additionally, relations with humans and between agents \citep{Fraune2020}, and related concepts like trust, empathy, rapport or presence have been extensively studied in multiparty hybrid interactions \citep{Gillet2022}, and have also shown their great potential to impact positively interactions. 

\citet{Traeger2020} illustrate how a robot's expression of vulnerability can enrich communication in a collaborative game, affecting not only human-robot relationships but also interhuman interactions. This phenomenon highlights the importance of intimacy, which relies on honesty, authenticity, and mutual understanding, in forming a healthy and engaging group dynamic. \citet{prager1995guilford} emphasizes that intimacy incorporates affective and cognitive components, involving emotional investment and an openness to others that can be crucial in collaborative interactions. Although intimacy has traditionally been studied in dyadic interactions \citep{Baur2013, Soleymani2019, Potdevin2021}, it offers an interpersonal framework where individuals can experience mutual understanding and a positive perception of themselves and others, thus contributing to the improvement of interpersonal relationships \citep{Potdevin2020}. This approach is relevant for exploring how socially interactive agents can enhance collaboration and cohesion within groups. 

The physical embodiment of interactive agents plays a crucial role in social interactions, influencing how these entities are perceived and engaged with by humans. Research has shown that physically embodied robots tend to elicit more favorable social responses compared to virtual agents. Li's  meta-analysis \citep{Li2015} found that physical robots outperform virtual agents in 73\% of cases studied, suggesting that their physical presence significantly enhances social interaction. This advantage is attributed to their co-location with users, which enhances engagement and anthropomorphism. \citet{Kiesler2008} supports this by demonstrating that physical embodiment increases salience and importance, leading to more meaningful and socially rich interactions. Participants in their study were more engaged and anthropomorphized a co-located robot more than a virtual agent. \citet{Mutlu2020} highlights how physical and virtual embodiments create different mental frameworks in users, affecting situativity, interactivity, agentivity, proxemics, and believability.

These differences are critical for understanding intimacy in human-agent interactions. Proxemics, the management of interpersonal space, is pivotal in regulating intimacy. Physical robots, sharing the same physical space, adhere to social and spatial norms, while virtual agents, constrained to digital environments, provide a safer space for emotional expression. This dynamic influence user perceptions and the extent to which they disclose personal information.

In light of this information, a pertinent research question arises: How does group composition in terms of embodiment  (two embodied conversational agents, two robots, or a robot and an embodied conversational agent, all expressing intimacy with two humans) affect group dynamics, perceived intimacy by humans, and the sense of belonging to the same group in collaborative tasks?

To address this main question, several sub-questions are explored:
\begin{enumerate}
    \item How does the presence of physical robots versus virtual agents influence group cohesion and dynamics in a collaborative context?
    \item How do different group compositions affect the perceived intimacy among human members?
    \item What impact does each group composition have on the humans' sense of belonging to the group?
\end{enumerate}

These questions aim to deepen the understanding of the mechanisms through which the embodiment of agents influences social interactions and performance in collaborative tasks. In this study, we propose an experimental protocol to study the role of embodiment on the perception of the level of intimacy displayed by agents. This study is part of a broader work, in which we want to investigate the skills that impact the social aspect of cohesion, and therefore, the skills relevant to build and strengthen interpersonal relations in a multiparty collaborative context. 

We chose to use a collaborative board game, "Mot Malin", as a use case for our work. In a first study, we collected a corpus of humans playing the "Mot Malin" game to analyze the patterns of interaction. This analysis enables us not only to specify the verbal and non-verbal behaviors of digital entities based on these data, but also to establish a baseline for comparing how the group's composition in terms of embodiment affects group dynamics and the sense of belonging.

We describe in the next section our collected corpus of human-human video interactions. Next, we describe our proposed experimental protocol including agents with different embodiment. Finally, we propose elements of discussion on our research questions and protocol.

\section{Material}

\subsection{Mot Malin : a collaborative board game}

\begin{figure*}[ht]
  \centering
  \includegraphics[width=0.8\linewidth]{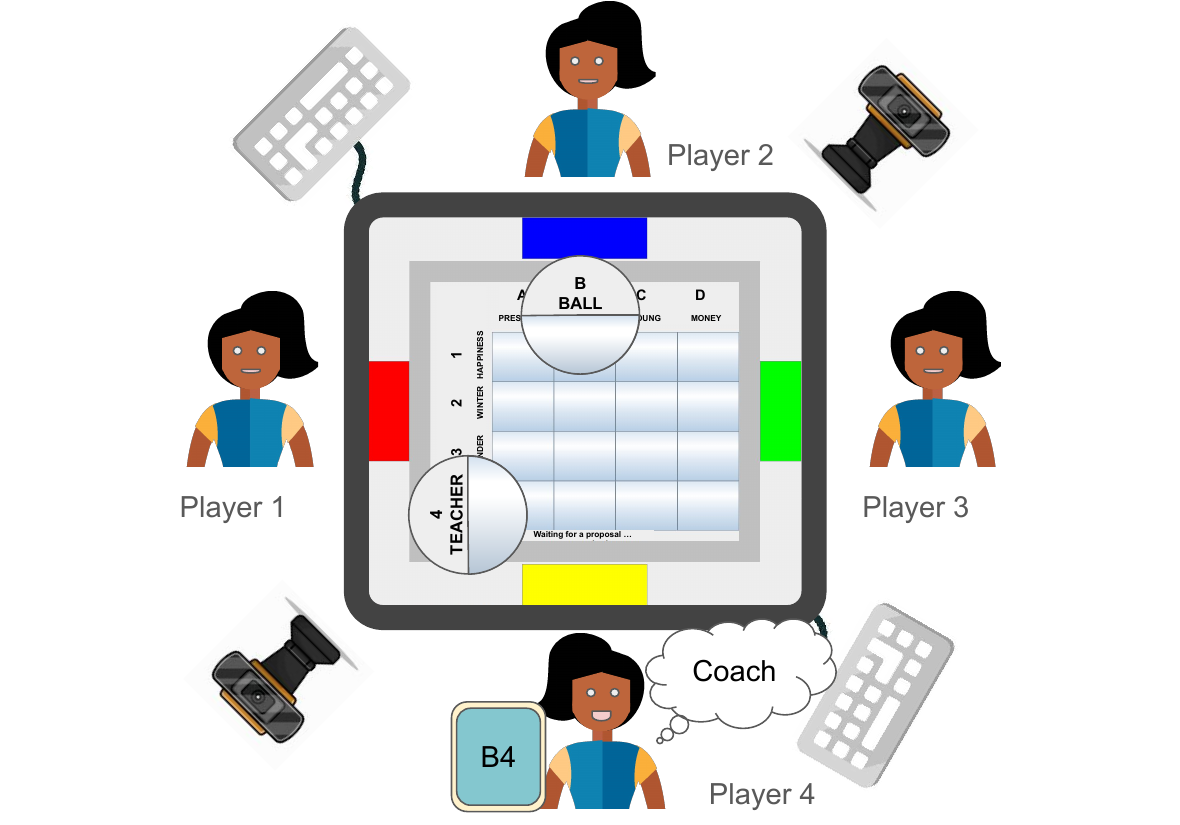}
  \caption{Experimental setup used to observe groups of 4 humans playing the game Computer version of the "Mot Malin" board game. The graphical user interface is displayed on a tactile screen in the center of a table. The players are seated around the table, equidistant from the screen, with two keyboards on opposite corner, each shared between two players. Two cameras on the other corners record the interaction.}
  \label{setup_mot_malin_hh}
\end{figure*}
 
We developed a computer version of the "Mot malin" board game. In this computerized version of the game, 4 players are seated around a table with a tactile screen in the center of the table, where the graphical user interface (GUI) is displayed, as illustrated in figure \ref{setup_mot_malin_hh}. Two cameras are placed near two opposite angles of the table, so that each camera captures two players at a time. The graphical interface is composed of 4 buttons on each side of the screen represented by colored rectangles. At the beginning of the session, a color (red, blue, green or yellow) is assigned to each player according to their position around the table. The game medium consists of a 4x4 grid displayed in the center of the GUI, where each column is identified with a letter (A,B,C,D) and associated with a word, and each row is identified with a number (1,2,3,4) and also associated with a word. All words are different, making a total of 8 different words. Every 16 possible coordinates on the board are therefore associated with a combination of a letter and a number that correspond to a pair of words. At the beginning of the game, each player receives a card with coordinates. Any player (one at a time) can decide to suggest a word related to the pair of words associated to the coordinates of their card by clicking on the button they are facing, before typing their proposition using one of the two keyboards. In the example (Figure \ref{setup_mot_malin_hh}), Player 4 has the card with the B4 coordinates. They decide to propose the word "coach" as it can be related to both "teacher" and "ball". The three other players have then to cooperate to guess the coordinate associated with the word suggested. They can click on the cell corresponding to their guess directly on the tactile screen to indicate their choice. When they all agree on a position, the first player has to announce if the proposition is correct and discard their card. If it is the case, the coordinate is marked as completed on the grid, otherwise nothing happens. The player who proposed the word receives a new card, and the game goes on until all 16 cards are played. 

\subsection{Experimental setting}
We recruited 6 groups of 4 humans (24 total, 1 female, 23 male, 0 other) aged between 21 and 45 (M=24.8 SD=5.57),  playing two rounds of the "Mot malin" game using the system presented in the previous section. We prepared two different grid for each round, presented in a random order. Each session was structured as follows : 
\begin{itemize}
    \item Consent : The experimenter presents the objective of the experiment and ask participants' consent to record and use the data collected. After describing the different steps of the experiment, the rules of the game are explained to the participants.
    \item Familiarization : Each participant takes a seat around the table, which remains unchanged during the whole experiment. During this phase, the GUI is introduced to the participants. The experimenters explain how to interact with it.    
    \item First game round : The experimenter starts the video recording.   
    \item Questionnaires : Once the round is over, the recording is stopped, and the participants fill in a set of questionnaires, presented in the next section.
    \item Second game round : same as the first round.
    \item Questionnaires : same set of questionnaires.
    \item Debriefing : The participants can give open suggestions about their feeling and the system.
\end{itemize}

\subsection{Measures}
After each game round, the participants were asked to fill in questionnaires to evaluate their perception of the interaction according to the following dimensions :         

\begin{itemize}
    \item Intimacy using the Virtual Intimacy Scale \citep{Potdevin2020} : Self report measure of the level of intimacy and measure of the level of intimacy of the facing participant. 
    
    \item Social skills : Evaluation of the facing participant's communication skills in multimodal social interactions using the Social Performance Rating Scale (SPRS) \citep{HametBagnou2022} adapted from \citep{Fydrich1998} and collaborative problem solving skills using the Social Skills of Collaboration (SSC) scale \citep{HametBagnou2022} adapted from \citep{Hesse2015}.
    
    \item The need to belong using the "Echelle du Sentiment d'Appartenance Sociale" (ESAS) \citep{Richer1996} : Measure of the feeling of group intimacy and acceptation in the group.
    
    \item Sociogram : Each participant chooses one player they would play again with, one they would not play again with, one they think would want to play with them and one they think would not want to play with them.
    
    \item Cohesion : Evaluation of the global cohesion of the team in one item. 
    
    \item The user experience of the system using the french adaptation of MeCue \citep{Lallemand2017}
\end{itemize}

\subsection{Data processing}
The corpus collected includes 12 game sessions composed of two videos (two players on each video) and a recording of the game board. We add to these videos the logged action on the GUI (request to speak, proposition of a word, proposition of a coordinates, success and failure), and the questionnaires answers. We also plan to add annotations of each participant's behaviors : 
\begin{itemize}
    \item Emotions : We developed a automatic emotion detection model based on the MobileNetv2 architecture and using the Fer2013Plus dataset to detect a set of basic emotions : neutral, happy, sad, angry, fearful, surprised, disgusted.
    \item Gaze behaviors : We developed a gaze detection system based on MediaPipe Face Mesh to detect the participant's gaze direction among five predefined targets : one for each of the 3 other participants, the board and looking away.   
    \item Gestures : Manual annotations of gestures associated with specific moments of the game : propose a word, propose a coordinate, think about a word, think about a coordinate, win, lose.
    \item Verbal transcriptions : Automatic transcription and diarization of the game sessions. 
\end{itemize}

These annotations will help us collect information about how humans play and communicate during the game according to their profile collected with the questionnaires. This will allow us in turn to define behaviors displayed by the agents in the next experiment.

\section{Experimental protocol}

\subsection{Setup}
\begin{figure*}[ht]
  \centering
  \includegraphics[width=\linewidth]{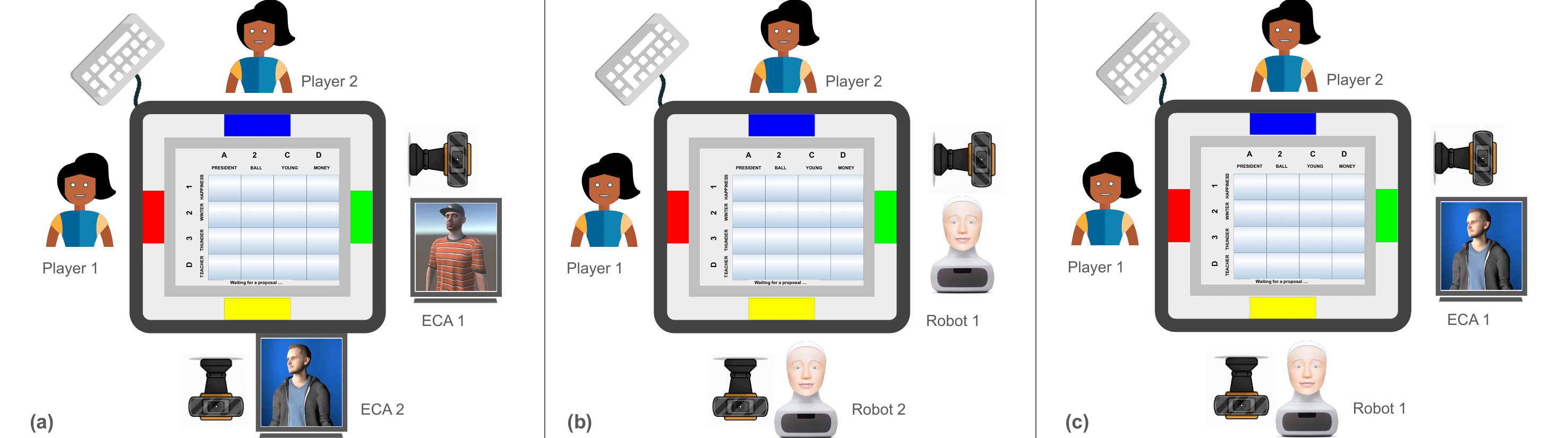}
  \caption{Illustration of the 3 experimental conditions of the next experimental protocol,  where 2 humans play the computer version of "Mot Malin" with : (a) ECA condition : 2 different looking ECA (b) Robot condition : 2 Furhat robots (c) Hybrid condition : 1 ECA and 1 Furhat robot. }
  \label{setup_mot_malin_hybride} 
\end{figure*}

In the next study, we will use the same experimental setup presented previously but with 2 humans and 2 agents with different embodiments. We define 3 experimental conditions as illustrated in Figure \ref{setup_mot_malin_hybride} : 
\begin{itemize}
    \item ECA (Figure \ref{setup_mot_malin_hybride}a) : The humans play with two Embodied Conversational Agents (ECA)  with different looks. The 3D models used can be animated (facial expressions, lips movements and gestures) through the Unity platform. The ECAs will be displayed on a vertically oriented screen, in a seated position, with only the upper body part visible to make it look like they are seated on the same table as the humans.

    \item Robot (Figure \ref{setup_mot_malin_hybride}b) : The humans play with  Furhat robots, which are humanoid robot heads with a projected virtual face that allows to fully animate facial expressions and lips movements, a speaker to play sounds, and a human-like neck structure allowing them to generate natural head movements \citep{AlMoubayed2012}.

    \item Hybrid (Figure \ref{setup_mot_malin_hybride}c) : The humans play with with one ECA and one robot.
\end{itemize}

In this setup, the two cameras will be placed behind the agents so that they can capture the human participants frontally, and only one keyboard, shared between the humans, will be used.  

\subsection{Measures}

Each group of human will play two game sessions in one of the three conditions, before evaluating the system using the same set of measures used during the first study for the intimacy, social skills, need to belong and sociogram measures. For the cohesion measure, we will replace the single item with a more complete questionnaire about cohesion, with the "Questionnaire sur l’Ambiance du Groupe" \citep{Heuze2002}, the Group Environment Questionnaire \citep{Carron1985} french adapataion, to have more insight on the group dynamic processes. We will exclude the user experience evaluation as it was mostly used to evaluate the GUI in the first study.

\subsection{Decision process and behavior generation}
The decision process related to (a) game play and (b) social behavior is the same for all agents. A finite-state machine related to the different steps of the game enable them to choose their next action accordingly. Depending on the situation (e.g. the right coordinate is found), they may select appropriate reactions towards the other players (e.g. joy).

As the different embodiments do not share the same features, these high-level actions are then translated by a dedicated module to low-level series of instructions, that include verbal and non-verbal behaviour. In the case of the Furhat robot, only facial expressions and head movements can be realized, while the ECA can perform gestures with its upper body.

The agents will display intimate behaviors towards other players (humans or agents), as it is shown that it has a positive impact on interpersonal relationships \citep{Short2017, Potdevin2021}. This will be reflected through non-verbal behaviors (e.g. smiles, head nods, eye contact or open hand gestures) and verbal behaviors (e.g. showing vulnerability by disclaiming weaknesses or making references to previously used words during the round) identified in the corpus and in the litterature.

\section{Discussion}
We presented in this paper a corpus collected in a previous study to analyze humans behaviors playing a computer adaptation of "Mot Malin", a collaborative board game. 

In its current state, the small size of the corpus and the unbalanced distribution of participants' gender and age does not allow statistically relevant observation patterns. Thus, we are still expanding the corpus by recruiting more participants with more diverse backgrounds. Another element of discussion concerns the questionnaires, and specifically the evaluation of intimacy and social skills. To avoid survey fatigue, we made the choice to ask the participants to only evaluate the participant in front of them for these scales. As a result, we lose the information of their evaluation of the other participants. For instance, the keyboards were shared between adjacent participants. We can ask ourselves if the fact of sharing a physical object, and what is more an interaction element in the game, could impact perceived intimacy and more globally group dynamics. 

This leads to a similar question about the next study, where we plan to use one keyboard shared between the human participants, which could lead the human participants to feel closer, and thus more intimate between each other than with the agents. We also did not consider already existing interpersonal relations between participants, while this can impact the group dynamics. For example, if three participants already know each other, this could lead to an increased sense of belonging between them, while the fourth player could feel isolated. The intial type of relationship between the agents should then be carefully defined, while taking note of how well the participants know each other. Moreover, we should take into account subgroups formation and their influence on the group dynamics.

We decided in the protocol to use ECAs with different looks, at the same level of detail, to ensure that the ECAs are perceived as two different entities. Similar agents might be systematically identified as an already bonded group by the humans participants, for example two ECAs with the exact same looks could be perceived as twins, which is commonly associated with complicity and connivence and could eventually make humans feel excluded. As the projected face of the Furhat robot can be customized by providing a flattened face image texture, we plan to use the same faces used on the ECA to avoid any bias on the looks and focus on the embodiment. In the hybrid condition, the faces will be randomized for each group. For the voices, we also plan to use the same methodology by using two distinct voices. Another important aspect is about the gender and ethnicity of the agents. While we are concerned about equity problematics in the IVA community, we decided to use agents with the same gender and ethnicity to avoid known biases and focus on the effects of embodiement.

From a technical point of view, some challenges remain. The first one concerns the ability of the agents to play the game itself, and how the agents will decide, according to the current state of the grid : (a) what word could be associated to the pair of words corresponding to their coordinate, and (b) what pair of words, or what coordinates, could match with the word suggested from another player. 

While it is time consuming, it is possible to predetermine one and even two possible words for each pair of words for the two grids (total of 32 to 64 words), but it is more complex to think about all possible words that could be suggested for each coordinate. We have studied automatic solutions for both cases, using different techniques, knowingly a lexical database, using WOLF, a french equivalent of WordNet, word embeddings using Word2vec, and Large Language Models (LLM), using Gemini. While we are still exploring the possibilities, the LLM solution seems to work best for the first issue, and word embeddings seems to give better results for the second case. 

But even if we know what word or coordinate to choose, another issue still exists, and it is to know which strategies should the agents use. For example, should they directly choose a coordinate after a word is proposed? Should they suggest to the other players only a component of the coordinate ? (e.g. "I think that the word coach is for sure related to teacher, but I am not sure about the other...") Should they lead the game and always suggest a word ? Or should they wait to be sure that no other player wants to suggest a word? We will draw inspiration from the corpus to choose a coherent and appropriate strategy for the agents according to the profile to apply to the agents. 

These last questions arise another discussion point, about turn taking, leadership, and the roles the agents should endorse during the game. Should they both have proactive or passive roles? Should they even have the same roles ? Is there a more adapted or accepted role for each embodiement? While no direct related measures is included in the experimental protocol about these concepts, existing works and observations on the future hybrid corpus we plan to collect could be investigated. We plan to define for the next study only a single role that both agents will endorse. 

Additional annotations on the corpus, like entitativity, synchrony or interruptions could also give us more in-depth understanding of the dynamics and give us leads to provide answers to these problematics.

\begin{acknowledgments}
This work was supported by a French government grant managed by the Agence Nationale de la Recherche as part of the France 2030 program, reference ANR-22-EXEN-0004 (PEPR eNSEMBLE / PC3 MATCHING).
\end{acknowledgments}

\bibliography{biblio}

\end{document}